\documentclass{icrc29}
\usepackage{graphicx,amssymb,amsmath,times}
\begin{document}
\title[Absolute calibration of imaging ...]{Absolute calibration of imaging atmospheric Cherenkov telescopes}
\author[N. Shepherd et al.] {N. Shepherd$^a$, J.H. Buckley$^b$, O. Celik$^c$, J. Holder$^d$, 
S. LeBohec$^a$, H. Manseri$^a$, F. Pizlo$^e$\newauthor and M. Roberts$^a$\\
  (a) Dept. of Physics, University of Utah, Salt-Lake-City, UT\\
  (b) Dept. of Physics, Washington University, St. Louis, MO\\
  (c) Dept. of Physics and Astronomy, University of California, Los Angeles, CA\\
  (d) School of Physics and Astronomy, University of Leeds, UK\\
  (e) Dept. of Physics, Purdue University, Lafayette, IN
}

\presenter{Presenter: S.LeBohec (lebohec@physics.utah.edu)} usa-shepherd-N-abs1-og27-oral

\maketitle

\begin{abstract}
A calibrated laser pulse propagating through the atmosphere produces a flash 
of Rayleigh scattered light with an intensity that can be 
calculated very accurately when atmospheric conditions are good. This is used 
in a technique developed for the absolute calibration of ultra high energy 
cosmic ray fluorescence telescopes, and it can also be 
applied to imaging atmospheric Cherenkov telescopes (IACTs). In this paper we present the 
absolute calibration system being constructed and tested for the VERITAS 
project.
\end{abstract}

\section{Introduction}
The absolute calibration of IACTs is usually
obtained from the lab measurement of the efficiency and gain of each detector element
(light collecting mirrors, photo-detectors, electronics, etc.). The
resulting calibration can be compared with muon arc images \cite{brian} 
although
complications may arise from the differences in the wavelength spectrum of the
light from muons and from the gamma ray showers of interest. Instrument performance 
is often monitored using relative calibration techniques
\cite{lebohec}.  The availability of a calibrated fast pulsed light source in 
the the sky would make the absolute calibration of these telescopes more
straightforward and reliable. Using the work done for atmospheric 
fluorescence detectors of ultra high energy cosmic rays \cite{Roberts03} 
\cite{Dawson02} as a guide, we have obtained a preliminary absolute calibration of the 
VERITAS telescope 1 by measuring the Rayleigh scattered light from 
a pulsed laser shot toward zenith. The technique relies on the fact that for a
known laser pulse energy and atmospheric temperature and pressure, the amount
of Rayleigh scattered light reaching the telescope to be calibrated can be 
computed with high accuracy and compared with the signal amplitude actually
recorded. 

\section{Observations}
The VERITAS experiment \cite{Weekes02} is an array of 
four 12 m telescopes under construction at Kitt Peak in Southern Arizona.  
Each telescope camera consists of 499 $\rm 0.15^o$ spaced photo-multipliers
arranged in a close packed hexagonal lattice covering a full field of view 
of $\rm 3.5^o$. The first telescope to be constructed has been temporarily
installed on the foothills of Mt-Hopkins at the base-camp of the Whipple 
observatory \cite{holder}. 

The calibration technique outlined above is complicated by fluctuating amounts
of aerosols present in 
the atmosphere. Because of Mie scattering, aerosols
increase the amount of light scattered off the laser beam. This is somewhat
compensated by the correspondingly increased atmospheric attenuation from the 
laser to the telescopes. As a result, there is a specific 
distance between the laser and the telescope for which the effects of aerosol 
fluctuations cancel out. This distance depends on the details of the aerosol
properties and has been shown to fall in the range of 3 to 5 km
\cite{Wiencke99} for horizontal observation. For the preliminary tests 
reported in this article, a nitrogen laser (337nm) pointed at zenith is 
installed 3.0~km away from the telescope. 

The telescope is pointed at the laser $\rm 20^o$ above the horizon  
and intercepts the laser when it reaches an altitude of 
$\rm \sim 1000m$ above ground level. Each laser pulse has a duration of 
4 ns and an average energy of $ \rm 25.5\mu J$ with a 2\% standard
deviation. The pulse remains within the field of view for $\rm \sim575~ns$ 
and moves across one $\rm 0.15^o$ pixel in $\rm \sim 25~ns$. The telescope 
is focused at infinity and the laser pulse produces a spot in the focal 
plane corresponding to $\rm \sim 0.22^o$. As a consequence, a given pixel 
can receive light from the laser pulse for as long as 60 ns. This also 
permits the laser flash to satisfy the trigger conditions, which require 
3 pixels to exceed their threshold within $\rm
\sim 7 ns$. 

Figure 1 shows the pixel map of one laser event. In these measurements we only
recorded 124 samples (248 ns) from the FADC which is why the laser beam 
image does not extend all across the entire field of view. The pulse width 
and relative timing are consistent with the geometrical considerations 
outlined above. The VERITAS system should allow the recording of signals 
for as long as $\rm 64\mu s$, so modifications to the acquisition software 
will permit longer integrations to be made in subsequent measurements. 

\begin{figure}[events]
\begin{center}
\includegraphics*[width=0.19\textwidth,angle=0,clip]{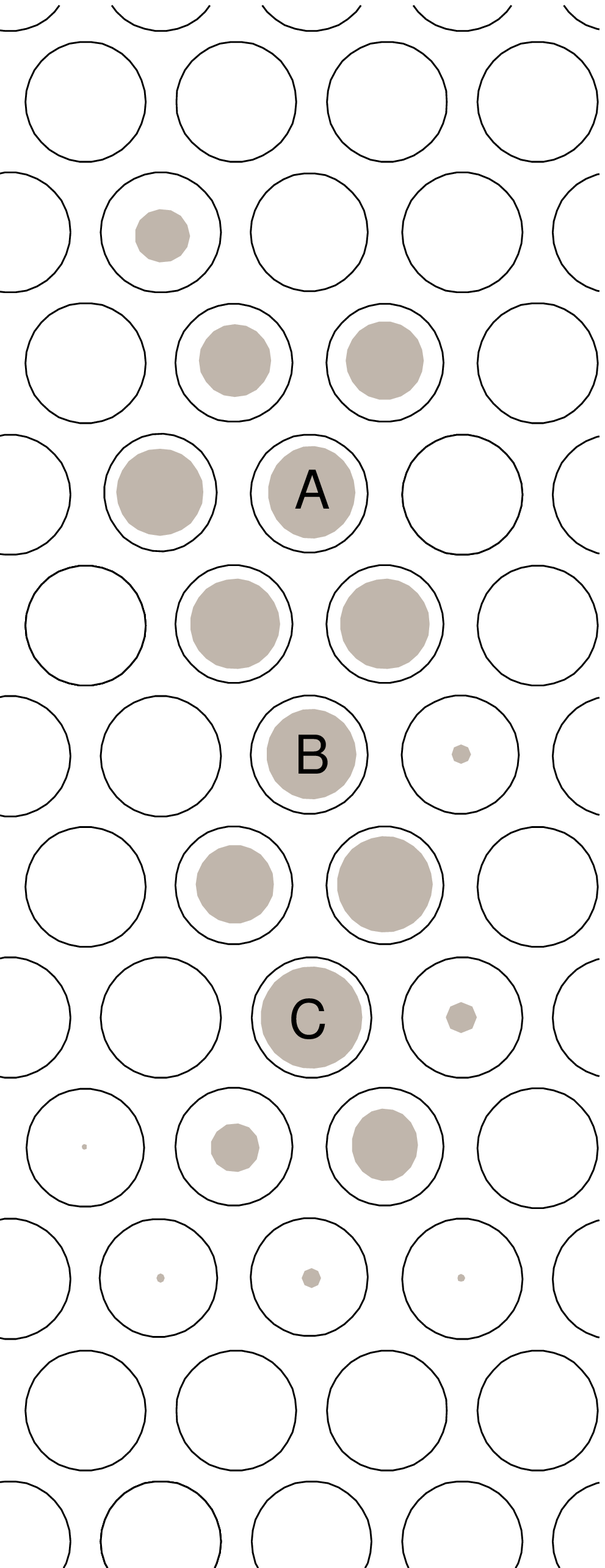}
\includegraphics*[width=0.7\textwidth,angle=0,clip]{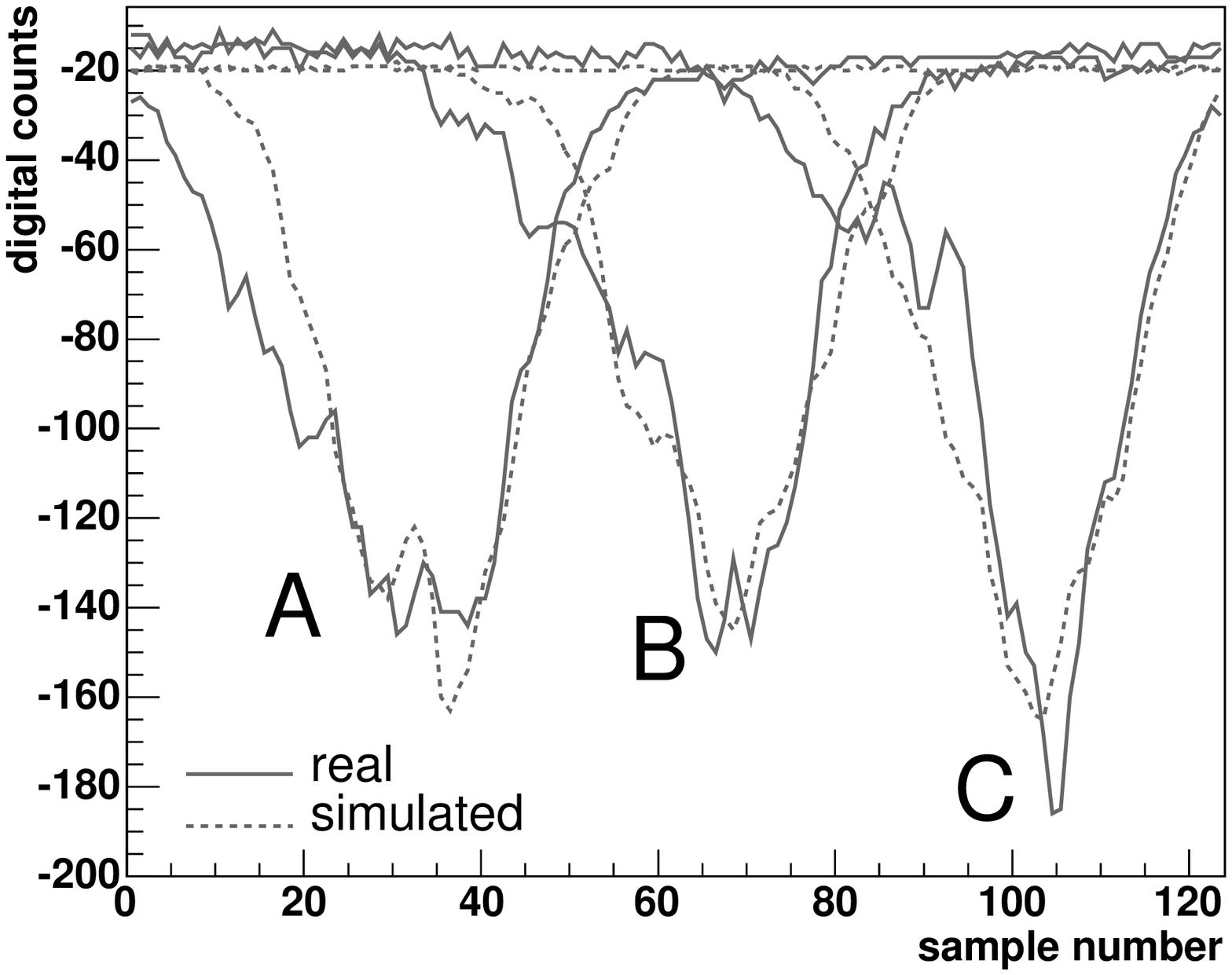}
\caption{\label {fig1} A laser event image is shown on the left. Each circle 
symbolizes one camera pixel the area of the grey disks indicate the time 
integrated signal from that pixel. The FADC traces are shown on the right for 
three pixels both in a real event (solid line) and in a simulated event 
(dotted line). Each sample corresponds to 2 nanoseconds.} 
\end{center}
\end{figure}

\section{Analysis}
In order to analyze the data obtained with the telescope, we developed a
detailed simulation of the experiment. The Rayleigh scattered light 
\cite{Bucholtz95} is simulated according to the measured laser pulse energy
($\rm 25.52 \mu J$), local temperature ($\rm 290.36^o K$) and pressure 
($\rm 88700 Pa$) and taking into account the geometry of the
setup. For this preliminary analysis, we have assumed the atmosphere to be
locally isothermal. The telescope response to the flash of Rayleigh scattered 
light was obtained by using the GrISU simulation package \cite{grisu}.
In the simulation, we positioned the laser to reproduce the event shown in
figure 1. We selected 39 channels for which we calculated the average pulse 
integral both from 29 real events and from 29 simulated events. The simulated 
events appeared to contain less signal than the real events by 
$12.0 \pm 2.9\%$ where the error is statistical only. This result can be 
verified on a pixel by pixel basis as in Figure 2. Although this 
figure illustrates the accuracy of our simulations, it should not be used to 
obtain an absolute calibration for individual channels since the contents of 
each pixel depends strongly on the precise position of the laser beam 
image in the field. In fact, figure 2 results from optimizing the position of 
the simulated laser beam image to obtain the strongest correlation between
simulated and real pixel contents. The resulting similarity of
simulated and real data can be further verified by comparing pulse shapes 
as in Figure 1. 

The systematic error is dominated by the residual
effects from Mie scattering on aerosols. This is further complicated in our 
observation by the observing angle, which makes us sensitive to the
aerosol vertical distribution. Systematic limitations can be obtained
experimentally and optimized by observing the night to night fluctuations and
minimizing them by adjusting the distance from the laser to the telescope. 

\begin{figure}[pixdata]
\begin{center}
\includegraphics*[width=0.6\textwidth,angle=0,clip]{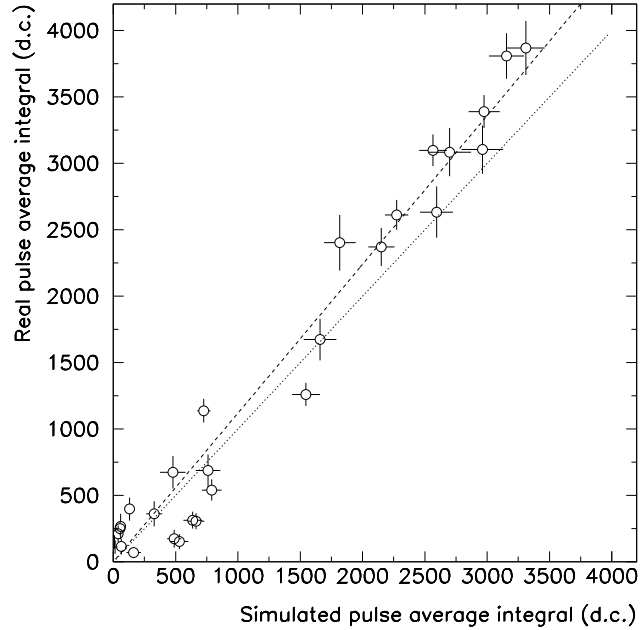}
\caption{\label {fig2} For each of the 39 selected pixels the averaged pulse
  integrals for 29 real events are shown as a function of the averaged pulse
  integrals for 29 simulated events. The dotted line represents equality
  between simulated and real data. The dashed line illustrates the 12\%
  discrepancy we observed.}
\end{center}
\end{figure}

\section{Conclusions}
The first results of this absolute calibration method applied to an imaging
Cherenkov telescope are very promising. Further improvement will come from the
elimination of the image time truncation, which complicates the
analysis. The local atmosphere model must be improved to allow for 
non-isothermal conditions. The effects of the distance from the laser to the 
telescope still have to be investigated for a better understanding of the 
systematics. 

\section{Acknowledgments}
This work was supported under US National Science Foundation Grant \# PHY
0099580. The authors wish to acknowledge their VERITAS colleagues
for their assistance during data collection and for their support 
with the analysis software.

\end{document}